\documentclass[11pt]{article}
\usepackage{blois,epsfig}

\def\Journal#1#2#3#4{{#1} {\bf #2}, #3 (#4)}

\def\PRL{\em Phys. Rev. Lett.}
\def\PRD{{\em Phys. Rev.} D}

\def\be{\begin{equation}}
\def\ee{\end{equation}}
\def\bea{\begin{eqnarray}}
\def\eea{\end{eqnarray}}

\newcommand{\beq}{\begin{equation}}
\newcommand{\eeq}{\end{equation}}

\begin{document}
\vspace*{4cm}
\title{SELF-FORCE AND MOTION OF STARS AROUND BLACK HOLES}

\author{ A.D.A.M. SPALLICCI DI FILOTTRANO}

\address{{D\'epartement de Physique et Sciences de l'Ing\'enieur, Universit\'e d'Orl\'eans}\\
{UMR 6115 Laboratoire de Physique et Chimie de l'Environnement et de l'Espace}\\
{UMS 3116 Observatoire des Sciences de l'Univers en R\'egion Centre}\\ 
{Campus CNRS, 3A Avenue de la Recherche Scientifique, 45071 Orl\'eans, France}\\
{spallicci@cnrs-orleans.fr}}

\author{ S. AOUDIA }

\address{{Observatoire de Paris, Section de Meudon}\\
{UMR 8102 Laboratoire Univers et Th\'eories}\\
{5 Place Julius Janssen, 92190 Meudon, France}}

\maketitle
\abstracts{
Through detection by low gravitational wave space interferometers, the capture of stars by supermassive black holes will constitute a giant step forward in the understanding of gravitation in strong field. The impact of the perturbations on the motion of the star is computed via the tail, the back-scattered part of the perturbations, or via a radiative Green function. In the former approach, the self-force acts upon the background geodesic, while in the latter, the geodesic is conceived in the total (background plus perturbations) field. Regularisations (mode-sum and Riemann-Hurwitz $\zeta$ function) intervene to cancel divergencies coming from the infinitesimal size of the particle. The non-adiabatic trajectories require the most sophisticated techniques for studying the evolution of the motion, like the self-consistent approach.}

\section{Introduction}

The relativistic two-body problem implies the emission of radiation and thus poses still formidable challenges even for radial fall \footnote{Back-action shows itself even in Newtonian physics, as the uniqueness of acceleration holds as long as the masses of the falling bodies are negligible. For free fall, the implications on the equivalence principle have been discussed.    
With the appearance of general relativity, the radial fall of a test particle into a Schwarzschild-Droste black hole has fueled a vivacious controversy in the first seventy years of general relativity but still echoing today on the existence of repulsion and on the velocity of a particle at the horizon \cite{ei82}.} \cite{sp10} and generally whenever adiabaticity can't be evoked. Indeed, adiabatic averaging intervenes if a sufficiently long period, in which energy-momentum balance may be applied, does exist. In curved spacetime, at any time the emitted radiation may backscatter off the spacetime curvature, and interact back with the particle later on: the instantaneous conservation of energy is not applicable and the momentary self-force acting on the particle depends on the particle's entire history \cite{quwa99}. Thus, the computation and the application of the back-action all along the trajectory and the continuous correction of the background geodesic, it is the only semi-analytic way to determine motion in non-adiabatic cases. Non-adiabatic gravitational waveforms are one of the original aims of the self-force community, since they express i) the physics closer to the black hole horizon ii) the most complex trajectories iii) the most tantalising theoretical questions.   

Perturbations were first dealt in 1957 \cite{rewh57}, when a Schwarzschild-Droste black hole was shown to regain stability after undergoing small vibrations about its spherical form, if subjected to a small perturbation. In the following forty years of analysis in the frequency domain, the captured mass radiates energy (the second time derivative of the quadrupole moment is different than zero), but its motion is still unaffected by the radiation emitted \cite{ze70a}. As analysis in the frequency domain is inherently limited, the breakthrough arrived thanks to a specifically tailored finite differences method, consisting of the numerical integration of the inhomogeneous wave equation in time domain \cite{lopr97b} and thanks to the determination of the self-force.

\section{The self-force}

It is only slightly more than a decade, that we possess methods for the evaluation in strong field of the self-force for point particles, thanks to concurring situations. On one hand, theorists progressed in understanding radiation reaction and obtained formal prescriptions for its determination and, on the other hand, the appearance of requirements from the LISA (Laser Interferometer Space Antenna) project for the detection of captures of stars by supermassive black holes (EMRI, Extreme Mass Ratio Inspiral), notoriously affected by radiation reaction. 

Before the appearance of the self-force equation and of the regularisation methods, the main theoretical unsolved problem was represented by the infinities of the perturbations at the position of the particle, represented by a Dirac delta distribution. 
In 1997, via the conservation of the total stress-energy tensor \cite{misata97}, or via the matched asymptotic expansion \cite{misata97}, or via the axiomatic approach \cite{quwa97}, but all yelding the same formal expression of the self-force, baptised MiSaTaQuWa from the surname first two initials of its discoverers, was found. On the footsteps of Dirac's definition of radiation reaction, in 2003, a fourth approach was presented \cite{dewh03}, hence MiSaTaQuWa-DeWh \cite{sp10}. A more rigorous way of deriving a gravitational self-force has been attempted without the step of Lorenz gauge relaxation \cite{grwa08}; an alternative approach and a new derivation of the self-force have been proposed \cite{gaspst06,po09}. A comprehensive living review online \cite{po04} and an upcoming book introduce the self-force (the entire volume previously cited \cite{sp10}).    

One pictorial description of the self-force refers to a particle that crosses the curved spacetime and thus generates gravitational waves. These waves are partly radiated to infinity (the instantaneous part) and partly scattered back by the black hole potential (the non-local part) thus forming tails which impinge on the particle and give origin to the self-force. 
Alternatively, the same phenomenon is described by an interaction particle-black hole generating a field which behaves as outgoing radiation in the wave-zone and thereby extracts energy from the particle. In the near-zone, the field acts on the particle and determines the self-force which impedes the particle to move on the geodesic of the background metric. 
The self-force is written as ($l$ represent the mode):

\beq
F_{self}^\alpha=\lim_{x\to z_u}\sum_{l}F_{tail}^{\alpha\,l}(x) = \lim_{x\to z_u}\sum_{l}\left[F_{full}^{\alpha\,l}(x)-F_{inst.}^{\alpha\,l}(x)\right]
\label{eq:diff-for-fself}
\eeq
where $F^\alpha_{full}$ represents the total contributions, $F^\alpha_{inst.}$ the contributions that propagate
along the past light cone, and $F^\alpha_{tail}$ the 
contributions from inside the past light cone, product of the
scattering of perturbations due to the motion of the particle in
the curved spacetime created by the black hole, fig.\ref{sf}. The self-force is then computed by taking the limit
$F^\mu_{self}=F^\mu_{tail}\left[x\to z_u(t)\right]$ ($x$ is the evaluation point, $z_u$ is the position of the particle on the worldline $\tau$). Thus, it may be conceived as force acting on the background geodesic \cite{misata97,quwa97}, in where $^0\!\Gamma^\alpha_{\beta\gamma}$ refers to the background metric:

\begin{figure}
\begin{center}
\includegraphics[width=1.9cm]{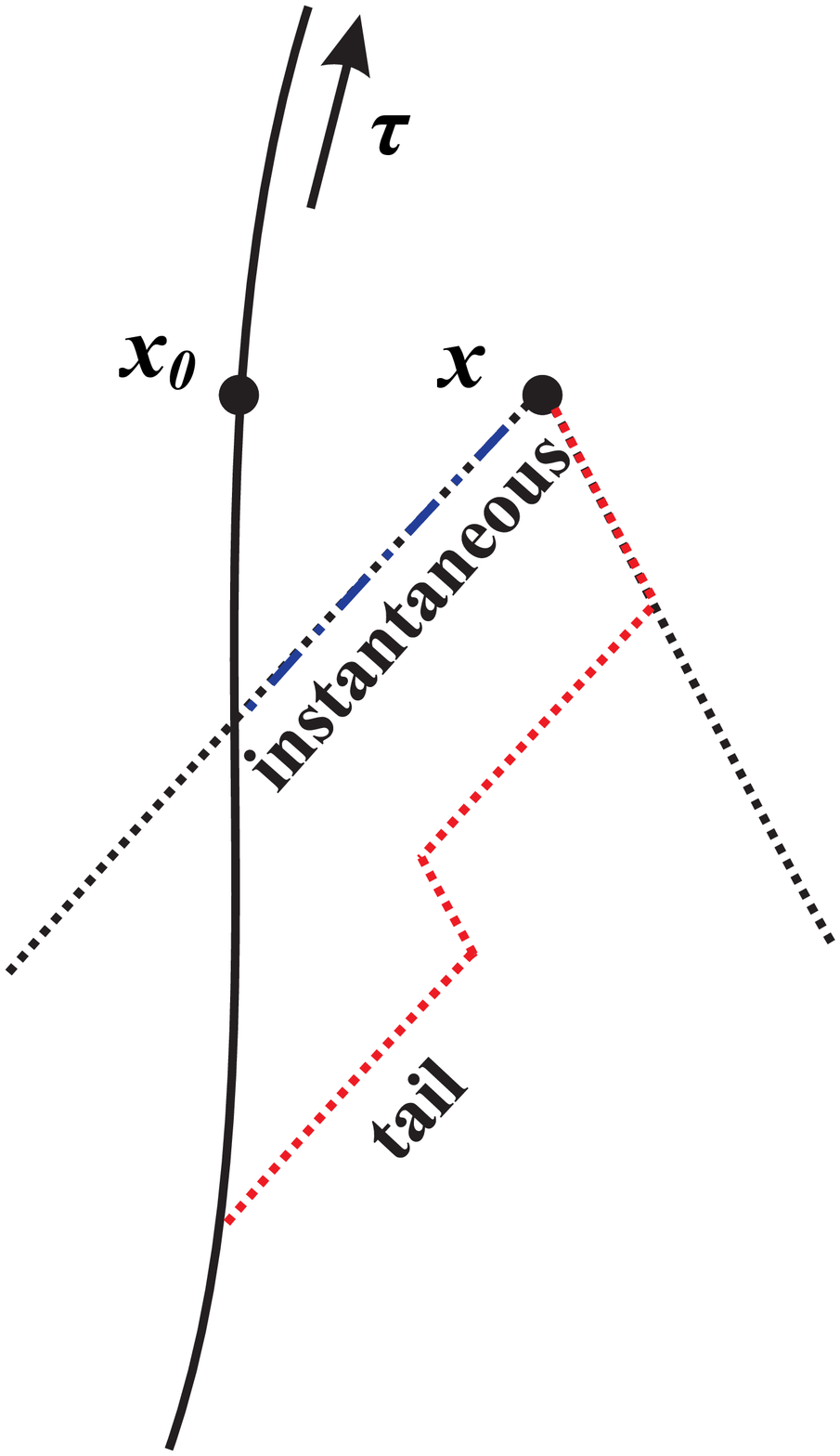}
\end{center}
\caption{The self-force, defined for $x$, evaluation point tending to $z_u$, particle position on the worldline $\tau$, is determined by the back-scattered radiation (tail). The instantaneous part of the perturbations goes to infinity.} 
\label{sf}
\end{figure}

\beq
F^\alpha_{self} =  m
\frac{Du^\alpha}{d\tau}= \frac{d^2 x^\alpha}{d\tau ^2}+^0\!\!\Gamma^\alpha_{\beta
\gamma} u^\beta u^\gamma 
\label{eq:sfformal}
\eeq
The MiSaTaQuWa-DeWh self-force acceleration is given by: 

\beq
a^\alpha = - (g^{\alpha\beta} + u^\alpha u^\beta)\left(\nabla_\delta
h_{\beta\gamma}^* - \frac{1}{2} \nabla_\beta h_{\gamma\delta}^*\right)u^\gamma u^\delta 
\label{eq:asf}
\eeq
where the star indicates the tail (MiSaTaQuWa) or radiative (DeWh) component. The self-force is only defined in the harmonic or Lorenz gauge and thus eq. (\ref{eq:asf}) is not gauge invariant and depends upon the Lorenz gauge condition:

\beq
{\bar h}^{\mu\nu\,*}_{;\nu}=0
\eeq
where ${\bar h}_{\gamma\delta}^* = h_{\gamma\delta}^* - 1/2 g_{\gamma\delta}h^*$ and $h^*=g^{\mu\nu}h_{\mu\nu}^*$. It has been shown \cite{baor01} that under 
a coordinate transformation of the form $x^\alpha \leftrightarrow x^\alpha + \xi^\alpha$, the particle acceleration changes according to 
$a^\alpha \leftrightarrow a^\alpha + a[\xi]^\alpha$ where the latter is the gauge acceleration.
Thus, for a given two-body system, the MiSaTaQuWa-DeWh acceleration is to be mentioned together with the chosen gauge \footnote{Being the self-force affected by gauge choice, the equivalence principle allows to find a gauge where the self-force disappears. Again, as in Newtonian physics, such gauge will be dependent of the mass $m$, impeding the unicity of acceleration.}. It should not be overlooked that the self-force is computed in proper time.  

The identification of the tail and instantaneous parts was not accompanied by a prescription of cancelation of divergencies, which indeed arrived three years later thanks to the mode-sum method \cite{baor00}. 
Spherical symmetry around a Schwarzschild-Droste black hole allows the force to be expanded into spherical harmonics. The divergent nature of the problem is then transformed into a summation problem. It has been shown that the self-force has an $1/l^2$ behaviour. 

An alternative pragmatic approach \cite{lo00,spao04} is the direct implementation of the geodesic in the full metric (background + perturbations) in coordinate time and it is coupled to a regularisation by the Riemann-Hurwitz $\zeta$ function. Though the application of the $\zeta$ function is somewhat artificial and the pragmatic method is somewhat naive, the latter has the merit of: i) a clear identification of the different factors 
participating in the motion; ii) potential applicability to any gauge of the $\zeta$ function regularisation and to higher orders. 
It has been shown \cite{balo02} the concordance of the mode-sum and $\zeta$ regularisations for radial fall. 

\section{Beyond the state of the art: the self-consistent prescription} 

For the evolution of an orbit, rather than a first order perturbation equation, containing geodesic deviation terms, a self-consistent approach has been recommended \cite{grwa08}. Such prescription affirms the greater accuracy of a first order perturbation development along a continuously corrected trajectory as opposed to a higher order perturbation development made on the background geodesic. Self-consistency bypasses the issue of relaxation, since at each integration step a new geodesic is found, and consists of a system of three 
equations to be simultaneously solved:

\beq
\nabla^\gamma \nabla_\gamma \tilde{h}_{\alpha\beta} - 2 R^\gamma{}_{\alpha\beta}{}^\delta \tilde{h}_{\gamma\delta} =
- 16 \pi M u_\alpha(t) u_\beta(t) \frac{\delta^{(3)}\left[x^\mu - z^\mu_p(t)\right ]}{\sqrt{-g}} \frac{d\tau}{dt}
\label{eq:sc1}
\eeq

\beq
u^\beta \nabla_\beta u^\alpha = - (g^{\alpha\beta} + u^\alpha u^\beta)(\nabla_\delta
h_{\beta\gamma}^{tail}- \frac{1}{2} \nabla_\beta h_{\gamma\delta}^{tail})u^\gamma u^\delta 
\label{eq:sc2}
\eeq

\beq
h_{\alpha\beta}^{tail}(x) = M
\int_{-\infty}^{\tau_{ret}^-}\left(G^+_{\alpha\beta \alpha '\beta '}-\frac{1}{2}g_{\alpha\beta}G^{+ \gamma}_{\gamma\alpha '\beta '}\right) \left[ x,z_p(\tau')\right ] u^{\alpha '}u^{\beta '} d\tau ' 
\label{eq:sc3}
\eeq
where $u^\alpha(\tau)$ in eq.s (\ref{eq:sc2},\ref{eq:sc3}), normalised in the background metric, refers to the self-consistent motion $z_p(\tau)$, rather than to a background geodesic; $G^+_{\alpha \beta \alpha '\beta '}$ is the retarded Green's function, normalised with a factor \cite{quwa97} of $-16 \pi$; the symbol $\tau_{ret}^-$ indicates the range of the integral being extended just short of the retarded time $\tau_{ret}$, so that only the interior part of the light-cone is used. Geodesic terms vanish in eq. (\ref{eq:sc2}), since self-consistency is imposed. 

\section{Conclusions}

The implementation of the self-consistent prescription is under consideration, but far from being gained. At each integration step, for a given number of modes, the process consists of: evaluation of the perturbation functions at the position of the particle; regularisation by mode-sum or $\zeta$ methods; correction of the geodesic and identification of the cell crossed by the particle; computation of the source term; reiteration of the above. Computationally and conceptually, it 
is a formidable challenge, but today one opportunity to study non-adiabatic motion for binaries having a small mass ratio.

\end{document}